# TWO COUPLED SUPERCONDUCTING CAVITIES AS A GRAVITATIONAL WAVE DETECTOR: FIRST EXPERIMENTAL RESULTS

Ph. Bernard, CERN, CH-1211, Geneva 23, Switzerland; G. Gemme[*] and R. Parodi, INFN, via Dodecaneso, 33, I-16146, Genova, Italy; E. Picasso, Scuola Normale Superiore, Piazza dei Cavalieri, 7, I-56126, Pisa, Italy

*Abstract*

First experimental results of a feasibility study of a gravitational wave detector based on two coupled superconducting cavities are presented. Basic physical principles underlying the detector behaviour and sensitivity limits are discussed. The detector layout is described in detail and its rf properties are showed. The limit sensitivity to small harmonic displacements at the detection frequency (around 1 MHz) is showed. The system performance as a potential g.w. detector is discussed and future developments are foreseen.

## 1 INTRODUCTION

The detection of weak forces acting on macroscopic bodies often entails the measurement of extremely small displacements of these bodies or their boundaries. This is the case in the attempts to detect gravitational waves (g.w.) or in the search for possible new long range interactions. In 1978 Bernard, Pegoraro, Picasso and Radicati suggested that superconducting coupled cavities could be used as a sensitive detector of gravitational effects, through the coupling of the gravitational wave with the electromagnetic field stored inside the cavities [1]. It has been shown that the principle that underlies this detector is analogous to the one used in parametric processes and in particular in frequency converters, i.e. in a device which converts energy from a reference frequency to a signal at a different frequency as a consequence of the time variation of a parameter of the system. In more detail the rf superconducting coupled cavities detector consists of an electromagnetic resonator, with two levels whose frequencies $\omega_s$ and $\omega_a$ are both much larger than the frequency $\Omega$ of the g.w. and satisfy the resonance condition $\omega_a - \omega_s = \Omega$. In the scheme suggested by Bernard et al. [1] the two levels are obtained by coupling two identical high frequency resonators. The frequency $\omega_s$ is the frequency of the level symmetrical in the fields of the two resonators, and $\omega_a$ is that of the antisymmetrical one. The g.w. trough its coupling to the electromagnetic energy can induce a transition between the two levels, provided their angular momenta along the direction *z* of propagation of the g.w. differ by 2. This can be achieved by putting the two resonators at right angle.

In 1984 Reece, Reiner and Melissinos (RRM), built a detector of the type proposed in the 10 GHz frequency range, and used it as a transducer of harmonic mechanical motion [2]. In order to measure the sensitivity limit of the detector, one of its walls was excited by an harmonic perturbation (by a piezoelectric); with a quality factor of the superconducting cavity equal to $3 \cdot 10^8$ and a stored electromagnetic energy approximately equal to 0.1 mJ they showed a sensitivity to relative deformations $\delta x/x \approx 10^{-17}$ (Hz)$^{-1/2}$. Since in the RRM experiment the interest was to measure small harmonic displacements and not gravitational effects, they used two identical cylindrical cavities mounted *end-to-end* and coupled via a small circular aperture in their common endwall. We wish to repeat the RRM experiment and improve its sensitivity by a factor $10^3$, thus reaching a sensitivity to harmonic displacements of the order $10^{-20}$ (Hz)$^{-1/2}$. If these goal would be obtained this detector could be an interesting candidate for the detection of gravitational waves or in the search of long range interactions of weak intensity.

## 2 BASIC PHYSICAL PRINCIPLES

A parametric converter is a nonlinear device which transfers energy from a reference frequency to a signal with different frequency, utilizing a nonlinear parameter (a reactance) of the system, or a parameter that can be varied as a function of time by applying a suitable signal. The time varying parameter may be electrical or mechanical; in the latter case the device acts as a transducer of mechanical displacements.

The basic equations describing the parametric converter are the Manley-Rowe relations [3]. They are a set of power conservation relations that are extremely useful in evaluating the performance of a parametric device. We will not derive here the complete Manley-Rowe relations, but we will just describe the fundamental ideas upon which the parametric converter behavior is based.

Let us consider a physical system which can exist in two distinct energy levels. To fix our ideas let us take as an example a system of two identical coupled resonant cavities like those of the RRM experiment. If the resonant frequency of the unperturbed single cell is $\omega_0$, then the frequency spectrum of the coupled system consists of two levels at frequencies $\omega_s$ and $\omega_a$, where $\omega_s=\omega_0-\delta\omega$ and $\omega_a=\omega_0+\delta\omega$, where $2\delta\omega/\omega_0=K$, is the coupling coefficient of the system.

---

[*] E-mail: Gianluca.Gemme@ge.infn.it

We can store some electromagnetic energy in the system at frequency $\omega_s$; if some external harmonic perturbation at frequency $\Omega = \omega_a-\omega_s$ induces the time variation of one system parameter we can have some energy transfer between the two energy levels, i.e. from $\omega_s$ to $\omega_a$. The external perturbation can equally well be an harmonic modulation of the cavity end wall, which causes the variation of the system reactance, or, in the case of the passage of a gravitational wave, the induced time modulation of the permittivity of the vacuum.

Let us call $P$ the total power absorbed by the device. We can write

$$P = P_s + P_a = \dot{N}_s \hbar\omega_s + \dot{N}_a \hbar\omega_a \quad (1)$$

where $\dot{N}_i$ is the time variation of the number of photons in level $i$. If we assume that the total number of photons in the systems $N = N_s + N_a$ is conserved we have

$$\dot{N}_a = -\dot{N}_s = \dot{N} \quad (2)$$

The total power absorbed by the system is therefore

$$P = \dot{N}\hbar(\omega_a - \omega_s) \quad (3)$$

and the power transferred to the upper level is given by

$$P_2 = \dot{N}\hbar\omega_a = \frac{P}{\hbar(\omega_a - \omega_s)}\hbar\omega_a = \frac{\omega_a}{\Omega}P \quad (4)$$

The amplification factor $\omega_a/\Omega$ is characteristic of parametric frequency converters.

To get deeper insight into eq. (4) we have to know the exact expression of $P$; in fact the power exchanged between the external perturbation and the system is proportional to the square of the fractional change of the time varying system parameter. If we denote this quantity with $h$, we can write for our coupled resonators system:

$$P = \Omega U_1 Q h^2 \quad (5)$$

From eq. (4) and eq. (5) we can easily derive the expression for the power transferred to the initially empty level

$$P_a = \omega_a U_s Q h^2 \quad (6)$$

In the last equation $U_s$ is the energy stored in level $s$, $Q$ is the electromagnetic quality factor of the resonant cavities.

The quantity $h$ above may represent the fractional change of the system reactance, or of the system length being harmonically modulated ($\delta x/x$), or the dimensionless amplitude of the g.w.

Since $P_a$ is proportional to the electromagnetic quality factor, superconducting resonant cavities should be employed to achieve maximum sensitivity.

## 3 SENSITIVITY LIMITS

Equation (6) could be a good starting point for a detailed discussion of the sensitivity limits of the detector. To make quantitative statements we need to know in some more detail our detector geometry. We choose a configuration very similar to the RRM experiment with two cylindrical niobium cavities coupled through a small circular aperture on the axis. The operating mode is the $TE_{011}$ at 3 GHz which, due to the coupling, splits into a symmetrical and an antisymmetrical mode respectively at frequency $\omega_s$ and $\omega_a$. The system is designed so that mode separation is about 1 MHz (see next section for more details on system design). For our geometry the relation between the maximum energy that can be stored in a superconducting niobium cavity and the frequency of the electromagnetic field is (in the following were not differently specified we take $\omega_s \approx \omega_a = \omega$)

$$\omega^3 U_{max} \approx 8.9 \cdot 10^{31} \text{Joule}\left(\frac{\text{rad}}{\text{sec}}\right)^3 \quad (7)$$

From eq. (7) we find at 3 GHz:

$$\omega U_{max} = \frac{8.9 \cdot 10^{31}}{\omega^2} \approx 2.5 \cdot 10^{11} \text{Joule}\frac{\text{rad}}{\text{sec}} \quad (8)$$

Deriving $h$ from eq. (6) and using the above result we get:

$$h_{min} \approx \left(\frac{P_a^{min}}{Q\omega U_{max}}\right)^{1/2} (\text{Hz})^{-1/2} \approx 10^{-16}\omega\left(\frac{P_a^{min}}{Q}\right)^{1/2} (\text{Hz})^{-1/2} \quad (9)$$

where $P_a^{min}$ is the noise power spectral density in our system. From eq. (9) is apparent that to get better sensitivity lower frequencies and higher quality factors are preferred.

For example at 3 GHz with $Q = 10^{10}$ and setting a noise power spectral density $P_a^{min} = 10^{-22}$ Watt/Hz we get $h_{min} \approx 2 \cdot 10^{-22}$ (Hz)$^{-1/2}$; while at 300 MHz, with $Q = 10^{11}$ and the same noise power spectral density one should get $h_{min} \approx 6 \cdot 10^{-24}$ (Hz)$^{-1/2}$.

To give a more realistic estimate of $P_a^{min}$ we have to discuss the various noise contributions present in our apparatus.

### 3.1 Symmetrical mode leakage

To operate our device we have to feed microwave power into one resonant mode and then to perturb one system parameter at a frequency equal to the mode separation, in order to detect the energy transfer between the full and the empty mode. Here we suppose that the initially full mode is the symmetrical one, and that this same mode is the lower frequency one. This is very likely to be the case, but is not at all crucial for the following discussion.

To feed power into our device we shall use a voltage controlled microwave oscillator locked onto the cavity symmetrical mode, at frequency $\omega_s$. If we assume a lorentzian power distribution and a constant power of the lower (symmetrical) level we get:

$$P_{osc} \approx \frac{\frac{4P_0 Q}{\omega_s}}{1+Q^2\left(\frac{\omega}{\omega_s}-\frac{\omega_s}{\omega}\right)^2} \frac{\text{Watt}}{\text{Hz}} \quad (10)$$

Here and in the following equations the suffix $s$ labels quantities related to the symmetric mode, while the suffix $a$ stands for the antisymmetric mode. From the above equation we can estimate the local oscillator noise power spectral density at the antysimmetric frequency, i.e. the leakage of mode $s$ at the frequency of mode $a$:

$$P_a^{osc}(\omega_a) = \frac{\frac{4P_0 Q}{\omega_s}}{1+Q^2\left(\frac{\omega_a}{\omega_s}-\frac{\omega_s}{\omega_a}\right)^2} \approx R \cdot \frac{U_s}{Q^2}\left(\frac{\omega_a}{\Omega}\right)^2 \frac{\text{Watt}}{\text{Hz}} \quad (11)$$

where we used the approximate relation $P_0 Q \approx \omega_s U_s$ and put $\Omega = \omega_a - \omega_s$.

$R$ is a number which depends upon the details of the measurement; if the receiver does not discriminate the parity of the field at frequency $\omega_a$, $R$ is of order one. Discrimination of the parity, as in the RRM experiment, would considerably reduce this value. RRM made use of a magic tee to excite the symmetric mode trough the $\Sigma$ port, and to detect the antisymmetric mode trough the $\Delta$ port (see fig. 1a). With careful adjustments they obtained 70 dB isolation between the $\Sigma$ and $\Delta$ ports of the magic tee ($R \approx 10^{-7}$).

Additional noise suppression should be achieved by improving the mode discrimination and detecting $P_a$ in pure transmission (see fig. 1b). With the use of two additional cavity couplings and another magic tee, the input and output signals could be optimally separated. The input couplings would be balanced so as to null the excitation of the second cavity mode produced by phase noise in the input. The output couplings would likewise be balanced to null the transmission of the symmetrical mode. By this arrangement the 70 dB isolation given by a balanced magic tee is used twice. In fact numerical simulations performed on this system configuration gave $R \approx 10^{-13}$ for the transmission detection scheme. With this calculated value of $R$, setting $U_s = U_{max}$ and $Q = 10^{10}$ we get, at 3 GHz, $P_a^{osc} \approx 10^{-19}$ Watt/Hz for $\Omega = 1$ KHz. At $\Omega = 1$ MHz, the contribution of the symmetrical mode width is negligible compared to the thermal noise spectral density: $P^{thermal} \approx kT = 2.5 \cdot 10^{-23}$ Watt/Hz at $T = 1.8$ K.

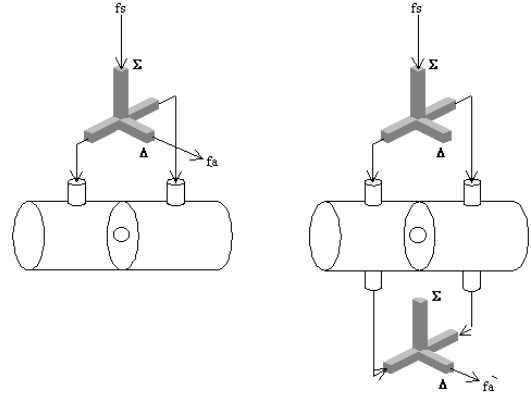

Figure 1: Schematic view of the experimental detection scheme in reflection and in transmission.

### 3.2 TWT Input Noise

The rf signal at $\omega_s$ feeding the cavity is generated by a VCO and amplified by a TWT amplifier before entering the first magic tee. The TWT amplifier is a Logimetrics A340/S with $A = 42$ dB nominal gain at 3 GHz and a noise figure $N_f = 18.7$ dB. The noise power spectral density input in the system is given by

$$P_a^{amp} \approx kT \cdot 10^{N_f/10} \cdot 10^{A/10} = kT_{eq} \approx 5 \cdot 10^{-15} \text{ Watt/Hz} \quad (12)$$

at 300 K. This value is, for the time being, the most severe limitation to system sensitivity, as we shall discuss in section 3.4. This huge noise power should be attenuated by the mode discrimination system described in the previous section. In fact the antisymmetric mode component excited by the TWT noise at frequency $\omega_a$ should be rejected by a factor of order $R$ by the input magic tee. First measurements performed on the system showed that, while the symmetric mode amplitude is attenuated by a factor of $10^{-8}$ by the discrimination system, this noise spectrum is virtually unaffected, being lowered, at most, by a factor of 10. This shows that the transmission detection scheme, while very promising, requires very careful adjustments of *both* the input and output couplings to be effective.

### 3.2 Low Noise Amplifier Input Noise

The input Johnson noise of the first amplifier in the detection electronics has to be evaluted and added to the former to establish the overall system intrinsic noise level.

The rf amplifier we are using is a JCA23-4029, manufactured by JCA Technology, Inc. which provides 48 dB gain at 3 GHz with a noise figure $N_f = 0.6$ dB.

This leads to a thermal noise power spectral density at room temperature equal to

$$P_a^{amp} \approx kT \cdot 10^{N_f/10} = kT_{eq} \approx 5 \cdot 10^{-21} \text{ Watt/Hz} \quad (13)$$

We point out that this value is obtained at $T = 300$ K which corresponds to $T_{eq} \approx 340$ K. In principle using a cryogenic preamplifier one should gain more than an order of magnitude and get $P_a^{amp} \approx 2.8 \cdot 10^{-22}$ Watt/Hz with an equivalent temperature $T_{eq} = 20$ K.

## 3.3 Brownian motion noise

Since we plan to operate our device as a mechanical motion transducer we must consider mechanical noise sources as well as electronic noise sources. In fact using eq. (6) we can write the brownian noise spectral density as

$$P_a^{Br} \approx \omega_a U_s Q \left(h^{Br}\right)^2 \quad (14)$$

To derive an explicit expression of $h^{Br}$ we have to develop a mechanical model of our system. Since it is quite difficult to treat the problem in the general case, and since we are interested in an order of magnitude estimate of the relevant quantities, we shall focus our attention on the longitudinal vibrational behaviour (i.e. on the mechanical displacements that change the cavity length) in two limiting cases:

- a low frequency limit $\left(\omega_m \approx \dfrac{\pi c_s}{L}\right)$, where the one-dimensional vibrational spectrum can be described by a discrete set of isolated resonances;

- an high frequency limit $\left(\omega_m >> \dfrac{\pi c_s}{L}\right)$, where the system can be considered an elastic continuum.

### 3.3.1 Low frequency limit

The mean square displacement spectral density near a mechanical resonance is given by [4]

$$|x(\omega)|^2 = \frac{\dfrac{4kT\omega_m}{mQ_m}}{(\omega^2 - \omega_m^2)^2 + \left(\dfrac{\omega\omega_m}{Q_m}\right)^2} \frac{m^2}{Hz} \quad (15)$$

This amplitude is, from the parametric conversion process point of view, equivalent to the external perturbing signal. In other words we can associate to brownian motion a fractional change of the characteristic system length $L$

$$h^{Br} \approx \frac{1}{L}\sqrt{|x(\omega)|^2} \approx \frac{1}{L}\left(\frac{\dfrac{4kT\omega_m}{mQ_m}}{(\omega^2 - \omega_m^2)^2 + \left(\dfrac{\omega\omega_m}{Q_m}\right)^2}\right)^{1/2} \frac{1}{\sqrt{Hz}} \quad (16)$$

Obviously the relevant contribution of brownian motion noise is for $\omega \approx \Omega$. We can slightly simplify the former expression in two limiting cases. When $\Omega \approx \omega_m$ we have

$$\left(h^{Br}\right)^2 \approx \frac{4kTQ_m}{mL^2}\frac{1}{\Omega^3}(Hz)^{-1} \quad (17)$$

while if $\Omega >> \omega_m$ we have

$$\left(h^{Br}\right)^2 \approx \frac{4kT}{mL^2}\frac{1}{Q_m}\frac{\omega_m}{\Omega^4}(Hz)^{-1} \quad (18)$$

From eqs. (14), (17) and (18) we can write for the brownian noise power spectral density

$$P_a^{Br} \approx \frac{4kT}{mL^2}Q_m Q \frac{\omega_a}{\Omega^3}U \quad (19)$$

for $\Omega \approx \omega_m$, and

$$P_a^{Br} \approx \frac{4kT}{mL^2}\frac{Q}{Q_m}\frac{\omega_a \omega_m}{\Omega^4}U \quad (20)$$

for $\Omega >> \omega_m$.

To give a numerical estimate of this quantity the mechanical properties of the resonator should be known. If we set $T = 1.8$ K, $m = 10$ Kg, $L = 1$ m, $Q_m = 100$, $Q = 10^{10}$ and $U = U_{max}$, we get near a resonance for $\Omega \approx \omega_m = 1$ KHz, $P_a^{Br} \approx 10^{-11}$ Watt/Hz. From the above considerations is easily seen that to get an high sensitivity, in this experimantal configuration, we should avoid to work at frequencies near to the detector mechanical resonances.

### 3.3.2 High frequency limit

Let us start our analysis by asking which is the mean vibrational energy per unit bandwidth in our system. We know that the mean energy per resonant mode is $\varepsilon = kT$, where we can safely neglect quanto-mechanical corrections for all frequencies of interest ($\hbar\Omega << kT$). To find out the mean energy per unit bandwidth we have to multiply the mean energy per mode by the number of modes per unit bandwidth at frequency $\Omega$: $\sigma(\Omega)$. It is well known that for a one-dimensional system this number does not depend on frequency and is equal to $\sigma(\Omega) = \dfrac{L}{\pi c_s}$, where $c_s$ is the sound velocity in the solid considered. We have for the mean vibrational energy per unit bandwidth:

$$\left\langle\frac{dE}{d\Omega}\right\rangle = kT\frac{L}{\pi c_s} \quad (21)$$

If we compare the above expression with the energy of a one-dimensional harmonic oscillator at frequency $\Omega$ we can write $\langle dE \rangle = m\Omega^2 \langle x^2 \rangle d\Omega$, where $\langle x^2 \rangle$ is the mean square displacement spectral density and we can easily derive an expression for $\langle x^2 \rangle$:

$$\langle x^2 \rangle = \frac{kT}{m\Omega^2} \frac{L}{\pi c_s} \quad (22)$$

At this point is straightforward to find the expression that gives $h^{Br}$ as a function of system parameters:

$$(h^{Br})^2 = \frac{\langle x^2 \rangle}{L^2} = \frac{1}{\pi c_s} \frac{kT}{m\Omega^2 L} \quad (23)$$

Using eq. (23) and setting $T = 1.8$ K, $m = 10$ Kg, $L = 1$ m, $c_s = 4.75 \cdot 10^3$ m/sec, $Q = 10^{10}$ and $U = U_{max}$, we get for $\Omega = 1$ MHz, $P_a^{Br} \approx 10^{-20}$ Watt/Hz.

## 3.4 System sensitivity

All of the noise sources are incoherent. The total noise power spectral density is the sum of all the noise sources: $P_a^{min} = P_a^{osc} + P_a^{TWT} + P_a^{amp} + P_a^{Br}$.

Using the expressions found above we find the behaviour shown in fig. 2 of system sensitivity as a function of detection frequency. The values plotted in fig. 2 have been calculated for a signal to noise ratio of 1.7, corresponding to 90% confidence level for gaussian probability distribution. Due to the strong dependence of detector sensitivity to the mechanical properties of the device a detailed study of those properties is needed.

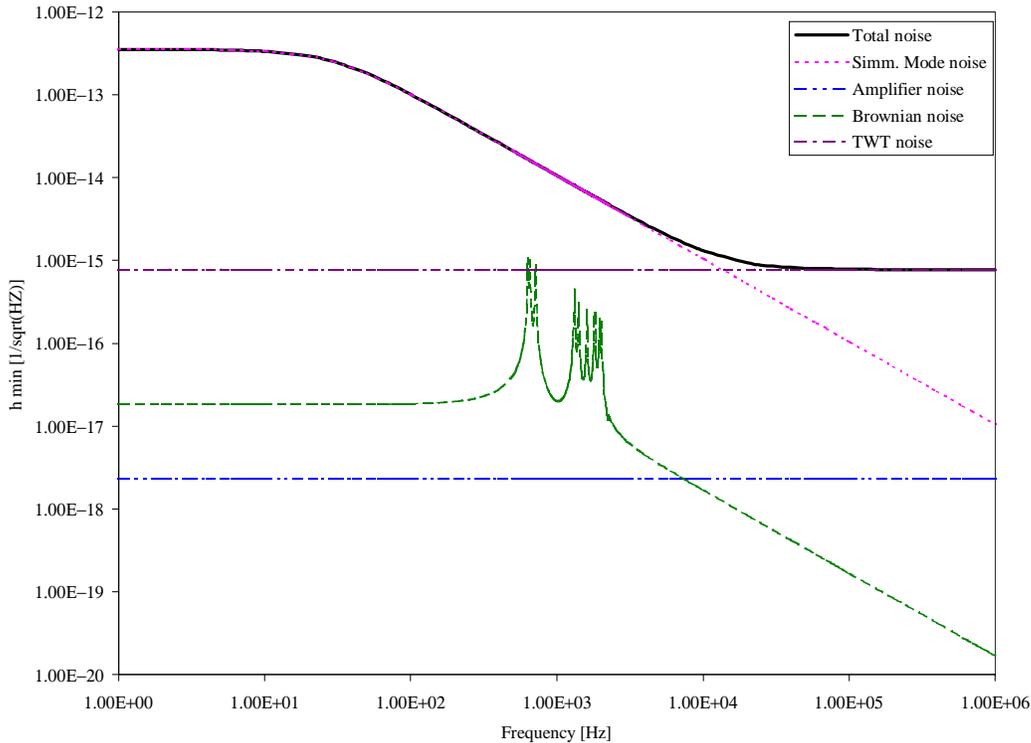

Figure 2: System sensitivity vs. detection frequency. Plotted values are for a signal to noise ratio of 1.7 (90% confidence level)

## 4 MECHANICAL AND ELECTROMAGNETIC DESIGN OF THE DETECTOR

The detector is built using two coupled RF cavities. The geometry of the resonators gives us the electromagnetic field needed for the detection of the wanted physical quantities.

Because the ultimate sensitivity of the detector at a given frequency is related to the $Q$ and the stored energy of the resonator, a resonator geometry with high geometric magnetic factor $\Gamma$ is preferred, where

$$\Gamma = \omega_0 \mu_0 \frac{\int_V H^2 dV}{\int_S H^2 dS} \quad (24)$$

To avoid rf electronic vacuum discharges and Fowler-Nordheim like non resonant electron loading, rf modes with vanishing electric field at the surface are mandatory.

The aforementioned requirements on the field configuration in the coupled resonator force us to follow the same path of Reece, Reiner and Melissinos choosing a TE mode for the cavities.

Early analysis of the best resonator shape suggested to use two spherical cavities coupled trough an iris [5];

nonetheless the higher cost of this choice and the dimension of the spherical resonator suggested us to use as a first step a conservative and lower cost approach using two cylindrical cavities coupled trough an axial iris.

The choice of the frequency, at this stage, was imposed by the maximum dimension for the resonator that can be housed in our standard test cryostat in a comfortable way; in our case the inner diameter is 300 mm giving us enough room for a 3 GHz resonator.

The naive choice to start our design is a couple of cylindrical resonators having the height $h$ equal to the diameter $d$.

For this configuration we can compute all the relevant quantities, as geometric factor and frequencies, starting from analytical expressions for the fields, or using the Bethe [6] perturbation theory for the evaluation of the coupling coefficient and mode separation.

Using a pill box like $TE_{011}$ geometry nevertheless should be a little bit upsetting due to the degeneracy of the $TE_{01}$ and $TM_{11}$ modes; sure this problem is quite mild in our case: due to the very high $Q$ value of the superconducting cavities, any distortion of the cavity geometry will split the TE and TM modes avoiding unwanted interactions.

Nevertheless we would like to enhance the splitting to be sure to remove any possible interaction between the TE and TM resonance.

To do that we design the cavity with a little modified geometry substituting the straight end plates with a spherical segment; the effect of this modification is to move away by 50 MHz the $TM_{11}$ modes.

To compute the resonant frequency and the rf quantities relevant for our experiment we used our Oscar2D code [7].

The coupling iris is a circular hole on the cavity axis. The effect of the coupling gives a symmetric field distribution at the lower frequency and the anti symmetric at the higher frequency, i.e. the coupling is electrical.

On the basis of the results of the simulations the final design of the cavity was decided: the construction drawing of the PACO resonator is shown on figure 3.

Unfortunately, during the electron beam welding of the bottom plates on the cylindrical barrel, two holes were produced on the niobium sheet. This caused a severe degradation of the cavity surface and, as a consequence, of the quality factor. The detector, before being tested, was chemically polished at CERN with standard niobium recipes.

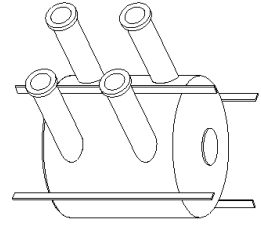

Figure 3: View of the final design of the detector.

## 5 EXPERIMENTAL RESULTS

The electromagnetic properties were measured in a vertical cryostat after careful tuning of the two cells frequencies.

The symmetric mode frequency was measured at 3.03431 GHz and the mode separation was 1.3817 MHz.

The unloaded $Q$ at 4.2 K was $5 \cdot 10^7$, and no significant improvement was found lowering the helium bath temperature at 1.8 K. Even after a second chemical polishing, performed at CERN, which removed approximately 300 μm of niobium from the surface, no improvement was observed. We believe that this very low $Q$ value is due to hot spots on the surface caused by the aforementioned welding problems. A second resonator, with slightly modified geometry, is under construction, and will be ready at the end of 1999.

Adjusting the phase and amplitude of the rf signal entering and leaving the cavity, the arms of the two magic tees were balanced to launch the symmetrical mode at the cavity input and to pick up the antisymmetrical one at the cavity output. With 30 dBm of power at the $\Sigma$ port of the first magic tee, -50 dbm were detected at the $\Delta$ port of the second one, giving an overall attenuation of the symmetric mode of 80 dB.

The energy stored in the cavity with 30 dBm input power was approximately 1 mJ.

The signal emerging from the $\Delta$ port of the output magic tee was amplified by the LNA and fed into a spectrum analyzer. The signal level at the antisymmetric mode frequency was –120 dBm in a 1 Hz bandwidth; the main contribution to this signal was the input noise of the TWT amplifier used to feed the cavity. Since the TWT noise passes trough the system with practically no attenuation and the rejection of the symmetric mode is relatively poor (four orders of magnitude worse than the expected value) the adjustments at the input and output port have to be improved.

System sensitivity at this stage is given by

$$h = \sqrt{\frac{P}{\omega U Q}} \approx 6 \cdot 10^{-16} \, (\text{Hz})^{-1/2} \qquad (25)$$

This value is quite far from our goal of $h \bullet 10^{-20} (\text{Hz})^{-1/2}$. Since, to our knowledge, this is the first example of a parametric detector operated in trasmission, and since this configuration requires very careful adjustments of the input and output ports balancing, we believe that significant improvements are obtainable. Furthermore the new cavity under construction should show the high quality factor needed to reach high sensitivity.

## 6 FUTURE PLANS AND CONCLUSIONS

We shall concentrate our attention to the following main items:
- System stability over time ranges of the order of magnitude of the inverse of our detection bandwidth. Obviously if we are going to detect our signal in a narrow bandwidth, we have to check carefully that system parameters do not change significantly over a time scale as long as possible. In particular the most critical parameter is the frequency splitting between the two normal modes $\Omega$. Experimental tests will show if an active control is needed to lock $\Omega$ at a fixed value.
- System sensitivity, by measuring noise level without any perturbing external signal. A careful analysis and comparison between theoretical predictions and experimental results on noise level will be done.
- Verification of parametric conversion, measuring the converted signal at different external perturbation amplitudes. This point includes optimum system tuning to get best signal to noise ratio.
- Mechanical modes distribution and properties measurement.

At the end of the experimental tests on the cylindrical detector, we shall build and test the spherical cavities detector configuration, which should give best performances in view of gravitational waves detection, due to the more favourable electromagnetic geometric factor. Moreover, using niobium sputtered copper cavities, we should take advantage of the very high quality factors obtained so far with this technique.

If the goal sensitivity would be obtained, the proposed detector could be an interesting candidate in the search of gravitational waves or long range interactions of weak intensity. In fact it is conceivable to push down the rf operating frequency in the 300 MHz range and the mode splitting in the 10 kHz range, thus making a series of similar detectors, working at different frequencies and/or mode separations, covering the $10^4 - 10^6$ Hz range. This frequency range is beyond the resonant bar and large bandwidth interferometers operating frequency so that the proposed detector could be useful to cover the high frequency region of the spectrum.

Recent works [8] focused their attention on the dection of stochastic g.w. sources and in particular of the relic g.w. background and pointed out that a relic background detected at high frequency would be unambiguously of cosmological origin. The detection of stochastic g.w. sources could be done by correlating two (or more) detectors put at a distance small compared to the g.w. wavelength (so that the signals could be correlated) but large enough to be sufficient to decorrelate local noises. With this experimental arrangement sistem sensitivity could be increased by several orders of magnitude [9] making possible the detection of very low signal levels.

## 7 ACKNOWLEDGEMENTS

We wish to thank Dr. S. Farinon of INFN Genova for the contribution given in the development and analysis of the detector mechanical model and dr. E. Chiaveri of CERN for the cavity chemical polishing.